# A New Approach for Steady-State Analysis of a Hybrid ac-dc Microgrid

Mohammad Mehdi Rezvani
Electrical and Computer Engineering Department,
Louisiana State University Baton Rouge, USA
mrezva2@lsu.edu

Shahab Mehraeen
Electrical and Computer Engineering Department,
Louisiana State University Baton Rouge, USA
smehraeen@lsu.edu

*Abstract*— The dc grid became more popular, by emerging the distributed generations (DGs). Despite this popularity, the dc grid is not yet widely used because the majority of loads in a power system are ac, which means the ac grid is still the dominant grid in the power system. Therefore, the concept of a hybrid ac-dc microgrid was emerged because of this contradiction. Hybrid ac-dc microgrid was introduced in order to exploit the benefits of both ac and dc microgrids. However, the combination of both ac and dc microgrids will add more complexity to the network. Because in all studies for hybrid ac-dc microgrid, such as steady-state analysis or dynamic study, two sets of equations should be considered and solved either separately or simultaneously, the solutions that were presented before. These solutions increase the time of simulation and operation.

In this paper, a novel procedure for steady-state analysis of a general hybrid ac-dc microgrid is proposed. In this technique, the dc microgrids will be transferred to the ac side by proposing two lemmas and then the whole grid will be analyzed as one ac network. It will be proved that not only the new ac grid has the same power flow result with the initial topology of the ac-dc microgrid, but also the simulation time of the proposed method is less than the other existing methods.

*Index Terms*—Hybrid Microgrid, Steady State Analysis, Fault Analysis,

## I. INTRODUCTION

Three phase ac power systems have been used for more than 100 years for their efficient transformation of power at different voltage levels and over long distances. Currently, more renewable power generation types are connected in the low-voltage side of ac distribution systems as Distributed Generation (DG). High penetration of DG units and their power/energy management has been a major concern for contemporary power systems [1-3]. Ac microgrids [1-9] have been proposed to simplify the connection of renewable power sources to conventional ac systems to overcome associated technical and economic challenges. In numerous literature, ac grids are analyzed for steady-state operation and stability based on the ac synchronous generators [10-12]. The proposed power flow algorithm in [13] exploits the frequency of islanded ac microgrid in the Newton-trust region (NTR). In [14] the optimal power flow problem is solved for an islanded ac microgrid by using particle swarm optimization. In [15] a power flow procedure is proposed based on the voltage of dc buses. Therefore, any change in the dc voltage affects the flow of power through the converters. The safe operation of ac grid is discussed in [16].

Since most of the power grids are still of ac type, ac microgrids are still predominantly used; however, they can be connected to dc zones creating hybrid ac-dc networks [17-20]. Thus, hybrid ac-dc circuits will be the next generation of the power grids. Simple ac-dc circuits comprise distribution laterals that stem out of ac-dc converters and feed dc loads or deliver power to ac circuits. High-voltage dc (HVDC) transmission lines are other simple hybrid ac-dc structures that connect the ac circuits to dc ones through a power line or series of lines. Analyses of such simpler structures are given in the past literature. Such analyses include load flow, fault analysis, dynamic stability, and control, etc. What all the available ac-dc circuit analysis methods have in common is the assumption that the dc/ac load/impedance is known and can be transferred from one circuit type (i.e., ac or dc) to another at the point of ac-dc coupling. This assumption mandates simple circuit topologies such as lateral configurations. A conventional method of analyzing ac-dc circuits is to merge both ac and dc circuits through bi-directional converters and establish a hybrid ac-dc network [21-23].

From the literature review, the main gap in studies over the ac-dc microgrids is lack of a general/unified steady-state model for hybrid networks. Previous research works regarding ac-dc grids focus on the dynamical behavior of hybrid networks and associated aspects like designing controllers and fault detection. On the other hands, the steady-state analysis in former studies are a sequential type and cannot be generalized to the entire hybrid network from the system level perspective. Moreover, in all references, the hybrid ac-dc microgrid was considered as two separate ac and dc grids which are connected to each other through an ac-dc converter. This is another weakness of these references because these studies cannot be implemented for an interconnected hybrid ac-dc microgrid, such as one which is depicted in Fig. 1. The proposed method in this paper aims to obtain a unified model of an interconnected hybrid ac-dc microgrids to be utilized in steady-state analysis such as power flow, faults studies, and stability assessment.

## II. FAULT ANALYSIS

In past decades, faults in ac power systems is comprehensively investigated. A fault is considered as a short circuit occurrence in the grid. Fault analysis can be classified into two categories: 1) Symmetrical faults and 2) Asymmetrical faults. In symmetrical faults, a three-phase short circuit occurs somewhere in the system and currents/voltages remain symmetrical. Therefore, the system can be analyzed in the form of a single line model. Asymmetrical faults include Single Line to Ground (SLG) Line-to-Line (L-L) and Double Line to-Ground (DLG). Due to the lack of symmetry in this category, a fault has to be investigated in a three-phase model.

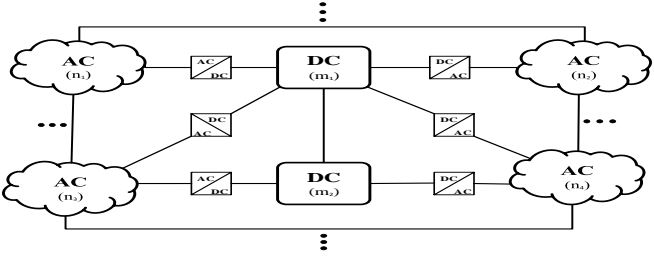

Fig. 1. A general model for an interconnected hybrid ac-dc microgrid

Prior to the fault analysis, loads and shunt capacitors can be neglected, as their current is significantly less than the fault current. This assumption leads to a voltage of $1\angle 0$ PU for all buses in the network. In an ac power system with *n* buses, if a symmetrical fault occurs at bus *k*, the steady-state bus voltages after the fault can be determined from (1) where the $Z_{bus}$ represents the inverse of network's $Y_{bus}$. Since $v_1^f$, …, $v_n^f$ are after-fault voltages, $v_k^f$ will be zero provided that the fault occurs at bus *k*. The fault current ($I_k^f$) can be straightforwardly determined by (2) where $Z_{kk}$ represents the element *kk* of the network's impedance matrix ($Z_{bus}$).

$$\begin{bmatrix} v_1^f \\ v_2^f \\ \vdots \\ v_k^f \\ \vdots \\ v_n^f \end{bmatrix} = \begin{bmatrix} 1\angle 0° \\ 1\angle 0° \\ 1\angle 0° \\ \vdots \\ 1\angle 0° \end{bmatrix} - [Z_{bus}] \begin{bmatrix} 0 \\ 0 \\ \vdots \\ I_k^f \\ \vdots \\ 0 \end{bmatrix} \quad (1)$$

$$I_k^f = \frac{1\angle 0°}{Z_{kk}} \quad (2)$$

Using (1) and (2), one can determine the fault current as well as the steady-state voltages affected by fault occurrence.

It can be seen from (1), that the $Z_{bus}$ matrix plays an important role in the fault analysis. This paper aims to develop a general method to construct the $Z_{bus}$ matrix for hybrid ac-dc networks so that the fault analysis would be based on a topology independent technique that can be generalized to all ac-dc networks regardless of the type of ac-dc interconnection.

The aforementioned equations describe the principle of classical fault analysis. Classical fault analysis is one of the major tools to analyze and design power systems. Transmission and distribution planning and operation of large-scale ac power systems rely highly on the simulation tools that perform such calculations. The classical fault analysis has several important applications including protection coordination and stability studies. Currently, available software tools utilize fault analysis efficiently but separately for ac circuits as well as dc ones; however, application of the available ac and dc analyses have not been extended to hybrid ac-dc circuits due to theoretical barriers.

Currently, a suite of analysis tools and techniques such as fault location using phasor measurement units (PMUs), network validation, load flow, fault current calculation, voltage compensation, time domain stability and swing analysis, classical control methods, etc., are prominently developed for ac power systems and to some extent for dc networks. The tools depend majorly on the model-based analysis of linear circuits using the admittance/impedance matrix. Through the classical fault analysis and using the admittance matrix, prediction of voltages and currents in all branches and nodes is possible. By contrast, the aforementioned tools cannot be easily extended to ac-dc hybrid power systems due to the existence of the conversion units in such circuits. That is, the effect of a fault in the dc circuit on the ac buses of an ac-dc hybrid network (and vice versa) cannot be predicted unless the iterative-based ac-dc load-flow is performed. The proposed unified ac-dc modeling makes ac-like linear circuit analyses possible for hybrid power networks through an equivalent ac admittance matrix and relaxes the time-consuming load-flow based fault analysis. This is a major improvement in speed of the ac-dc fault analysis. In the following, a simple hybrid ac-dc network is studied for steady-state fault analysis.

III. PROBLEM FORMULATION

In this paper, we are looking for an ac equivalent circuit that can represent the dc network that satisfies two conditions.

*Condition 1*

The RMS values of the dc-equivalent ac node voltages are equal to the magnitude of the corresponding dc node voltages in per unit.

*Condition 2*

The injected complex power at the points of ac-dc coupling in the equivalent ac circuit must be the same as those of the original ac-dc circuit while the active power at equivalent ac circuit is the same as dc powers at the corresponding nodes in the dc circuit.

*Lemma I*

*In order to satisfy Condition 1, all transferred dc voltages to the ac side of the circuit must be in-phase. In addition, all transferred dc currents to the ac side of the circuit must be in-phase.*

*Proof*

The proof can be conducted through contradiction. Consider the following circuit.

$$\begin{bmatrix} Y_1 & Y_2 \\ Y_3 & Y_4 \end{bmatrix} \begin{bmatrix} V_1 \\ V_2 \end{bmatrix} = \begin{bmatrix} I_1 \\ I_2 \end{bmatrix} \quad (3)$$

where $[V_1] = [v_1 \; v_2 \; \cdots \; v_m]^T$ and $[V_2] = [v_{m+1} \; v_{m+2} \; \cdots \; v_n]^T$ represent the voltages of those buses which are connected to ac-dc converters and those which are inside the dc grids, respectively. Therefore, the $I_1 = [I_1 \; I_2 \; \cdots \; I_m]^T$, which is the matrix of injected currents, is not zero and the $I_2 = [0 \; 0 \; \ldots \; 0]^T_{(n-m)*1}$. Here, n is the total number of dc buses, and m is the number of dc buses, which are connected to the ac-dc converters. The equations (4) and (5) would be obtained by manipulating in (3).

$$[Y_1][V_1] + [Y_2][V_2] = [I_1] \quad (4)$$
$$[Y_3][V_1] + [Y_4][V_2] = [I_2] = [0] \Rightarrow [V_2] = -[Y_4]^{-1}[Y_3][V_1] \quad (5)$$

$V_2$ will not give the true value of $v_2$ if $V_i$ (i=1, …, m), the elements of the matrix $[V_1]$, are not in-phase. Thus, $V_i$ (i=1, …, m) must be in-phase.

The equation (3) can be written such $YV = I$; where $Y$ is the admittance matrix; $V$ is the vector of reflected dc voltages to the ac circuit $V = [V_1 \ V_2 \ \cdots \ V_n]^T$, $I$ is the vector of reflected dc currents to the ac circuit $I = [I_1 \ I_2 \ \cdots \ I_n]^T$. The reflected dc values the ac side of the circuit are defined as $V_i = \alpha v_i \angle \theta$ where $\alpha = (\sqrt{3}/(2\sqrt{2}))M_a = 0.612 M_a$ and $v_i$ is the dc voltage at dc bus '$i$' and $I'_j = \beta i'_i \angle \gamma$ where $\beta = \frac{1}{\alpha}$ and $i'_i$ is the dc current injected to dc bus '$i$'. Where $M_a$ is the modulation factor of the ac-dc converters.

*Remark*

Here we consider all the reflected dc voltages and currents to the ac circuit are in-phase and thus angle $\theta$ and $\gamma$ are common among all the reflected ac nodal voltages and branch currents, respectively.

Using phase-shift transformer would be one possible solution to maintain these voltages in-phase. The model of this transformer is depicted in Fig. 2. The equation (6) represents the operation of this transformer. This transformer should be installed at the ac side of each ac-dc converter.

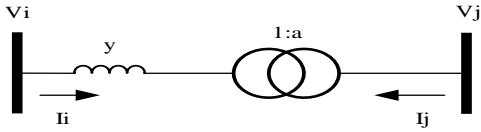
Fig. 2. Shift Transformer Model

$$\begin{bmatrix} y & \frac{-y}{a} \\ \frac{-y}{a^*} & \frac{y}{|a|^2} \end{bmatrix} \begin{bmatrix} V_i \\ V_j \end{bmatrix} = \begin{bmatrix} I_i \\ I_j \end{bmatrix}, a = |a|\angle \theta_a \quad (6)$$

Fig. 3 shows the configuration of the tested network. This configuration has been implemented in the MATLAB 2017b/SIMULINK to verify the proposed approach.

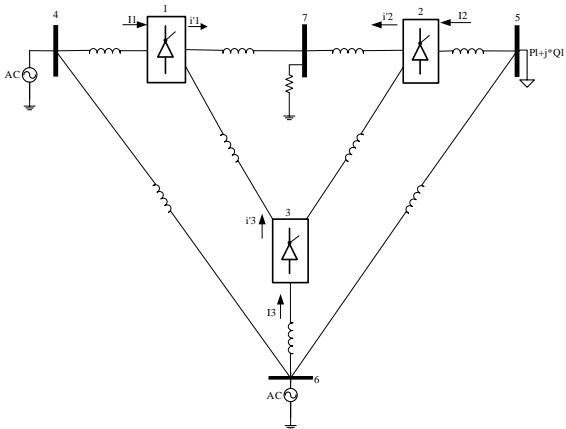
Fig. 3. Test ac-dc-ac hybrid microgrid

By taking into account the *lemma 1* for the hybrid ac-dc microgrid, depicted in Fig. 3

$$V_1 = |V_1|\angle\theta_1, \frac{V'_1}{V_1} = a_1 \xrightarrow{a_1 = 1\angle \theta_{a_1}} V_1 = |V'_1|\angle(\theta_1 - \theta_{a_1}) \quad (7)$$

$$V_2 = |V_2|\angle\theta_2, \frac{V'_2}{V_2} = a_2 \xrightarrow{a_2 = 1\angle \theta_{a_2}} V_2 = |V'_2|\angle(\theta_2 - \theta_{a_2}) \quad (8)$$

$$V_3 = |V_3|\angle\theta_3, \frac{V'_3}{V_3} = a_3 \xrightarrow{a_3 = 1\angle \theta_{a_3}} V_3 = |V'_3|\angle(\theta_3 - \theta_{a_3}) \quad (9)$$

Based on the *Lemma I*, all three phases must be equal which mean:

$$\angle(\theta_1 - \theta_{a_1}) = \angle(\theta_2 - \theta_{a_2}) = \angle(\theta_3 - \theta_{a_3}) \quad (10)$$

Since the dc network is an active grid, its transferred network should be active network too, which means it does not allow the reactive power to flow.

*Lemma II*

*In order to satisfy Condition 2, at least one impedance (capacitive and inductive element) should be installed at one of the dc buses.*

*$Y_{bus}$ construction*

Expressing a generic way to construct $Y_{bus}$ of the equivalent ac microgrid is the trickiest part for this problem. In this section, the $Y_{bus}$ of the equivalent ac microgrid is constructed based on the concept of Branch Building (BB) block. The BB block is a useful tool for constructing the $Y_{bus}$ between ac-side and dc-side of the converter.

Fig. 4 shows the typical topology of a converter node when both *lemma I* and *lemma II* are taken into account. Here, the shift transformer acts as an adjustment for both reflected dc-voltage and reflected dc current, and the current source acts as an impedance allowing the reactive power flow through the equivalent network. The vector diagram of this part of the network is illustrated in Fig. 5.

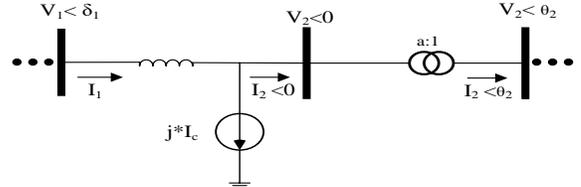
Fig. 4. The typical topology of a converter node

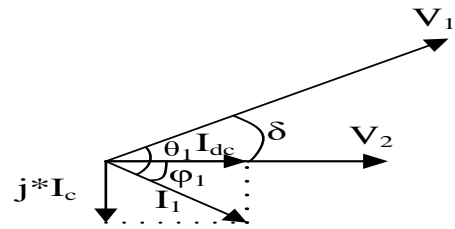
Fig. 5. Vector diagram of the Fig. 4

Based on the Fig. 4 the current $I_1$ would be obtained by (11).

$$I_1 = I_{dc} + jI_c = \frac{V_1 \angle \delta - V_2}{jX} = \frac{V_1 \cos(\delta) + jV_1 \sin(\delta) - V_2}{jX} = \frac{V_1 \sin(\delta)}{X} - j\frac{V_1 \cos(\delta) - V_2}{X} \quad (11)$$

Based on (11) the current $I_1$ consists of two parts, 1. the real part ($I_{dc}$), 2. the imaginary part ($I_c$). Therefore, the equation (11) could be split to two separate parts, which are needed to construct $Y_{bus}$.

$$I_{dc} + jI_c = \frac{V_1 \sin(\delta)}{X} - j\frac{V_1 \cos(\delta) - V_2}{X} \Rightarrow$$

$$I_{dc} \angle 0 = \frac{V_1 \sin(\delta)}{X} \Rightarrow V_1 = \frac{X \cdot I_{dc} \angle 0}{\sin(\delta)} \quad (12)$$

$$jI_c = -j\frac{V_1 \cos(\delta) - V_2}{X} = I_1 - I_{dc} \Rightarrow V_2 = \frac{X}{j}I_1 + \left(\frac{X}{\tan(\delta)} - \frac{X}{j}\right)I_{dc} \quad (13)$$

where $V_1$ is the voltage of real ac-node, $V_2$ is the voltage of reflected dc voltage after considering some phase shift in the angle.

The BB block based on the (12) and (13) is such as (14)

$$\begin{bmatrix} V_1 \\ V_2 \end{bmatrix} = \begin{bmatrix} 0 & \frac{X}{\sin(\delta)} \\ \frac{X}{j} & \left(\frac{X}{\tan(\delta)} - \frac{X}{j}\right) \end{bmatrix} \begin{bmatrix} I_1 \\ I_{dc} \angle 0 \end{bmatrix} \quad (14)$$

By multiplying the angle of $\angle \delta$ and $\angle \theta_2$ to (14), we have:

$$\begin{bmatrix} V_1 \angle \theta_1 \\ V_2 \angle \theta_2 \end{bmatrix} = \underbrace{\begin{bmatrix} 0 & \frac{X \angle \delta}{\sin(\delta)} \\ \frac{X}{j} & \left(\frac{X}{\tan(\delta)} - \frac{X}{j}\right) \end{bmatrix}}_{B} \begin{bmatrix} |I_1| \angle \varphi_I + \theta_2 \\ I_{dc} \angle \theta_2 \end{bmatrix} \quad (15)$$

and $\theta_1 = \delta + \theta_2$.

Finally, by multiplying the (15) by the inverse of matrix B the BB block will appear. By using the BB block, the $Y_{bus}$ matrix would be easily obtained.

$$\begin{bmatrix} |I_1| \angle \varphi_I + \theta_2 \\ I_{dc} \angle \theta_2 \end{bmatrix} = \underbrace{\begin{bmatrix} 0 & \frac{X \angle \delta}{\sin(\delta)} \\ \frac{X}{j} & \left(\frac{X}{\tan(\delta)} - \frac{X}{j}\right) \end{bmatrix}^{-1}}_{BB} \begin{bmatrix} V_1 \angle \theta_1 \\ V_2 \angle \theta_2 \end{bmatrix} \quad (16)$$

After unifying both ac and dc grids, the typical Newton-Raphson algorithm can be used for power flow calculation in order to steady-state studies. One of the most important benefits of this approach is, running power flow algorithm (Newton-Raphson method) only for one time, despite other methods which Newton-Raphson method was implemented for both dc and ac grid separately. Therefore, the proposed approach is faster than other methods.

## IV. CASE STUDY AND SIMULATION RESULTS

The goal of this part is to steady-state analysis of one interconnected hybrid ac-dc microgrid, which is depicted in Fig. 3, and validation of the proposed model.

The tested model consists of two ac voltage source, one ac load, one dc microgrid, three ac-dc converters with their filters, and three isolated generators.

It was assumed that a three-phase symmetrical fault occurs at 0.5 seconds and it removes at 0.83 seconds. As it is shown in Fig. 6, the steady state fault current during the fault period is 719.4 Amp.

The (16) is used to construct the BB block for each converter. Moreover, the angle δ in (16) is supposed to be 29°, 33.5°, and 29° for converters one, two, and three, respectively and it is considered to be constant during the simulation. These values are the initial voltages angles at the ac side of converters. The BB block for these three converters buses are such as follow:
For converters one and three:

$$BB = \begin{bmatrix} 3.215 - j5.8 & j6.6313 \\ 3.215 & 0 \end{bmatrix} \quad (17)$$

For converter two:

$$BB = \begin{bmatrix} 3.66 + j5.53 & j6.6313 \\ 3.66 & 0 \end{bmatrix} \quad (18)$$

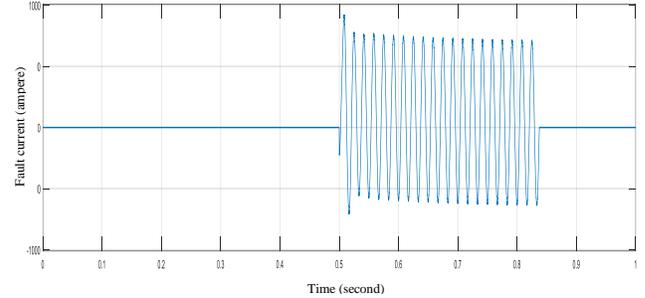

Fig. 6. The fault current

By constructing the BB block for other buses the $Y_{bus}$ and consequently the $Z_{bus}$ matrices of whole ac equivalent circuit can be easily obtained. The Thevenin impedance at fault bus is 0.0075 - j0.0774. Finally, the fault current will be calculated by dividing the voltage of faulty node by the Thevenin impedance, which is roughly 718.6104 Amp, which is roughly close to the MATLAB Simulink result (719.4 Amp).

## V. CONCLUSION

This paper presents a novel method for steady-state analysis of an interconnected hybrid ac-dc microgrid which is based on the ac equivalent circuit. The goal of this paper is to find the relation between dc microgrid and its reflection to the ac side in order to build a unit and comprehensive admittance matrix. Thus, the amount of computation will be significantly reduced because there is one set of equation that should be solved, which decrease the time and increase the speed of calculation.

To justify the proposed method the fault analysis has been covered for one interconnected hybrid ac-dc microgrid which is simulated in MATLAB Simulink. Then the result of MATLAB Simulink for fault current is compared to the result of the proposed method for the fault current of the specified hybrid grid. The results of MATLAB and the proposed method is a testimony to the fact that this method is valid.